# An Ensemble Deep Learning-based Cyber-Attack Detection in Industrial Control System

**Abdulrahman Al-Abassi[1], Hadis Karimipour[1], (Senior Member, IEEE), Ali Dehghantanha[1], (Senior Member, IEEE), Reza M. Parizi [2], (Senior Member, IEEE)**

[1]Cyber Science Lab, University of Guelph, Guelph, ON N1G 2W1 Canada
[2]College of Computing and Software Engineering, Kennesaw State University, GA 30060, USA

Corresponding author: R. M. Parizi (e-mail: rparizi1@kennesaw.edu).

**ABSTRACT** The integration of communication networks and the Internet of Things (IoT) in Industrial Control Systems (ICSs) increases their vulnerability towards cyber-attacks, causing devastating outcomes. Traditional Intrusion Detection Systems (IDSs), which are mainly developed to support Information Technology (IT) systems, count vastly on predefined models and are trained mostly on specific cyber-attacks. Besides, most IDSs do not consider the imbalanced nature of ICS datasets, thereby suffering from low accuracy and high false positive on real datasets. In this paper, we propose a deep representation learning model to construct new balanced representations of the imbalanced dataset. The new representations are fed into an ensemble deep learning attack detection model specifically designed for an ICS environment. The proposed attack detection model leverages Deep Neural Network (DNN) and Decision Tree (DT) classifiers to detect cyber-attacks from the new representations. The performance of the proposed model is evaluated based on 10-fold cross-validation on two real ICS datasets. The results show that the proposed method outperforms conventional classifiers, including Random Forest (RF), DNN, and AdaBoost, as well as recent existing models in the literature. The proposed approach is a generalized technique, which can be implemented in existing ICS infrastructures with minimum changes.

## I. INTRODUCTION

Critical infrastructures are highly complex systems that utilize cyber and physical components in their daily operations. The backbone of these facilities consists of an Industrial Control System (ICS), which plays an important role in the monitoring and control of critical infrastructures such as smart power grids, oil and gas, aerospace, and transportation [1] [2]. Therefore, the safety and security of ICSs are paramount for national security.

The inclusion of the Internet of Things (IoT) in ICSs opens up opportunities for cybercriminals to leverage the system vulnerabilities towards launching cyber-attacks [3] [4]. Awareness of the cyber-security vulnerability in ICSs has been growing since Stuxnet, the first cyber-attack that specifically targeted these technologies, revealed in 2010. Stuxnet intended to sabotage the system's operation without disturbing Information Technology (IT) systems [5]. In 2015, another cyberattack by the name of Black-Energy was used to target Ukraine's power grids, causing a massive power outage that affected about 230,000 people [6]. In February 2020, three U.S. gas pipeline firms announced another cyber-attack alleging a shutdown of electronic communication systems for multiple days [7]. While some of these attacks may result in information leakage, others can damage the physical system or misrepresent the system state to the monitoring engineer. These examples emphasize the growing cyber threat on Operational Technology (OT), which runs much of the enabling computer technologies that ICS in critical infrastructure (i.e., power, gas, and water), now rely on [2] [8].

While the security concerns of critical infrastructure facilities are already considered in the IT community, limited efforts have been made to develop security solutions that are specific to ICSs and OT environments [9]. Due to the differences between the nature and characteristics of IT and OT systems, these attacks mostly remain invisible to the traditional IT security measures such as Intrusion Detection Systems (IDSs) and anti-virus programs. Also, the communication protocols used by ICS (e.g., Modbus or DNP3 [10] and IEC standards [11]) are not adequately secured by traditional IDS. Therefore, strong security mechanisms are required to be designed explicitly for OT environments and ICSs to defend such attacks and to protect critical infrastructure facilities.

Different frameworks for IDSs have been used in the literature, such as model-based [12], and learning-based approaches [13] [14]. Most of these techniques utilize the available data to develop a model that exhibits the normal behavior of the system, then identify all different behaviors as abnormal. Since these methods are only trained on specific types of attacks, they are not able to detect unseen or new attack types [15] [16]. Besides, current IDSs are customized for specific systems/protocols, which lack adequate generalization [18].

Most importantly, the existing literature does not consider the imbalanced nature of ICS datasets, which results in low detection rates or high false positive in real scenarios [17]. A dataset is imbalanced if the instances of some classes are far fewer than other classes. The fundamental principle of classification is finding the boundary between different classes. If some classes are rarely presented, they may not be able to provide enough information to determine the boundary. Therefore, they may be treated as outliers resulting in wrong classifications.

Confronting these concerns, in this paper, we propose a generalized ensemble deep learning method for cyber-attack detection in ICS, which is evaluated on different real ICS datasets. The proposed deep learning model consists of multiple unsupervised Stacked Autoencoders (SAE) that learn new representations from imbalanced datasets. Then, new representations from each SAE are passed to a Deep Neural Network (DNN) via super vector and concatenated using a fusion activation vector. Finally, a Decision Tree (DT) is used, as a binary classifier, to detect attacks from the newly merged representations. Experiments show that the proposed model outperforms existing approaches with an acceptable performance even though fewer malicious instances are used.

The main contributions of the proposed method can be listed as follows:

- Developing a deep representation learning model to construct new balanced representations. The new representations increased attack detection accuracy and robustness (f-score) in an imbalanced environment.

- Increasing the detection accuracy and reducing the false positive rate by developing an ensemble deep learning algorithm based on DNN and DT classifiers to detect cyber-attacks from the new representations.

- Developing a generalized model that can be used in different critical infrastructure facilities with minimum changes in the existing system. The proposed framework utilizes representation learning and ensemble methods that can be trained to detect cyber-attacks in ICSs regardless of the data imbalance ratio.

The rest of this paper is structured as follows. Section II gives a literature review of recent studies in the field of ICS security. Section III presents a brief overview of the general ICS structure, system model, and different attack models considered in this work. The proposed method is described in Section VI. Section V includes results and case studies followed by the concluding remarks in Section VI.

## II. Related Work

Traditionally, ICSs were in an isolated environment with the focus on safety, where each system is safeguarded to stop the process if something goes wrong. However, the introduction of Internet protocols, IoT devices, and wireless technologies within ICSs has resulted in significantly less isolation from the outside world. Consequently, safety mechanisms, which were not designed to deal with malicious attacks, face more vulnerabilities than ever before.

The majority of current existing techniques on cyber-attack detection in ICSs are based on traditional IDSs, which are mainly designed for IT security analysis [5] [17]. IDSs can be categorized as signature-based and learning-based techniques. Signature-based approaches use databases and fixed signatures to detect known attacks, rendering them inefficient in detecting unknown or new attacks [19]. On the other hand, learning-based systems aim to identify process trends or behaviors that increase the efficiency to manage unexpected intrusions [20]. [21] used a common-path mining method for anomaly detection in smart cyber-physical grids. An attack detection technique based on the Pearson correlation between two sensor parameters was used in [22]. Authors in [23] utilized an IDS based on the Gaussian process to the attack strategy for anomaly detection. While these approaches are effective in detecting unusual activates, they are not reliable due to frequent upgrades in the network, resulting in different IDS topologies.

In contrast, learning-based IDSs are designed based on a moving target to continually evolve and learn new vulnerabilities [24] [25]. These methods try to generate the normal behavior of the system using existing datasets, then identify the irregular pattern as abnormalities. The authors of [26] proposed an anomaly detection technique based on reinforcement learning and convolutional autoencoders for ICS. Alternatively, [27] addresses the detection of Denial of Service (DoS) attacks using Support Vector Machine (SVM) and RF. [28] suggested an unsupervised technique for the effective detection of privacy attacks based on observations of eavesdropping attacks. [29] uses a variety of DNN methods, including different variants of convolutional and recurrent networks for cyber-attack detection in water treatment facilities. An ICS anomaly detection method using Long Short-term Memory (LSTM) networks is proposed in [30]. The authors of [31] proposed an attack detection techniques based on Hierarchical Neural Network. Similarly, [32] proposed a deep learning-based IDS through utilizing Recurrent Neural Networks (RNNs).

In another study [33], the authors applied a stacked Nonsymmetric Deep Autoencoder (NDAE) to develop their IDS. [34] proposed an unauthorized intrusion detection technique and conducted backdoor attacks on a SCADA Industrial Internet of Things (IIoT) testbed. [35] proposed a graphical model-based approach for detecting abnormal behavior in an ICS using Bayesian networks to map the relationship between sensors and actuators. [36] implemented a toolchain with multiple state-of-the-art Anomaly Detection (AD) techniques used for detecting attacks that appear as anomalies. Their findings suggest that detection rates can change dramatically when considering different detection modes, thereby necessitating a reliable and real-time AD technique to maintain resilience in critical infrastructures. [37] proposes a genetic algorithm (GA) to find the best NN architecture for a given dataset, using the NAB metric to determine the consistency and quality of different architectures. [38] evaluates the application of unsupervised machine learning algorithms, including DNN and SVM, to detect anomalies in the Cyber-Physical System (CPS) using data from a Secure Water Treatment (SWaT) testbed. Results indicate that the DNN classifier results in less false positives when compared to the one-class SVM, while SVM can detect more anomalies.

Although the above-mentioned works addressed some of the issues related to cyber-attack detection in ICSs, most of them are heavily reliant on feature engineering. These methods are quite complicated and require sophisticated learning techniques, which can potentially increase their computational burden. Furthermore, the majority of current proposed techniques are evaluated using balanced datasets, which lack the standard representation of imbalanced data in the ICS environment. Thus, it is hard to deploy such algorithms as they cannot extract various discriminative information from real-world imbalanced datasets. As such, in this paper, we propose a deep learning-based attack detection technique, which extracts a new representation from raw imbalanced datasets, for reliable and accurate attack detection with a low false-positive rate in highly imbalanced datasets from ICS environments.

## III. System Model

### A. Industrial Control Systems
A typical ICS network in a SCADA system architecture, as shown in Figure 1, consists mainly of a remote station, primary center, and regional center. These systems can interact with each other via wide/local area networks or Radio Telemetry. The primary center gathers data from field sensors, identifies new setpoints to track the operations of the network, and detects any existing irregularities. Then, instructions are sent to the remote station to monitor telemetry from field devices [39]. The regional station manages the network communication and regional power consumption between the primary and remote stations.

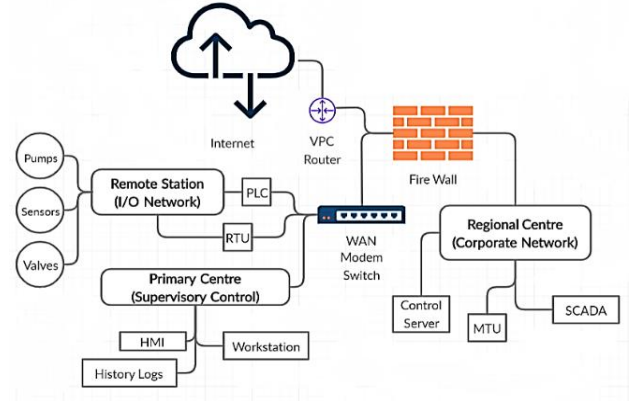

**FIGURE1.** ICS Standard Operation analysis model

ICS can be modeled using non-linear and non-Gaussian processes through the following equations:

$$x_k = g(x_{k-1}, \omega_k) \quad (1)$$
$$y_k = h(x_k, \upsilon_k) \quad (2)$$

where the state of the system is denoted by $x_k \in \mathbb{R}^n$ at time k. Sensor measurements are denoted by $y_k \in \mathbb{R}^m$. The process and sensor noise are denoted by $\omega_k$ and $\upsilon_k$ respectively.

### B. Adversary Model
The main attack types addressed in this study involve integrity attacks, such as False Data Injection (FDI) and availability attacks, such as DoS. In FDI attacks, an attacker executes the attack by injecting false data into the system shown in the equation below:

$$\tilde{y} = y + \alpha \, o \, y^a \quad (3)$$

where $\tilde{y}$ denotes the observation, $y$ is the true sensor measurement, $o$ is the element-wise multiplication, $y^a$ is the measurement noise, and $\alpha$ is the sensor-selection vector described below:

$$S_\alpha \triangleq \{\alpha \epsilon \mathbb{R}^m : \alpha_i = 0 \text{ or } 1, \sum_{i=1}^{m} \alpha_i = f\} \quad (4)$$

where the node $i$ is chosen as a malicious node and $\alpha_i$ is equal to 1. Typically, the intruder can exploit up to $f$ of $m$ sensors to fully inject false data into the system.

On the other hand, DoS attacks include measurement (packet) loss with two main types of modeling, including

Bernoulli distribution [40] and Markov model [41]. The attacker usually initiates DoS attacks by manipulating sensor readings and jamming communication channels, thereby flooding packets in the network [42]. This is illustrated below:

$$\mu_k(i) = \begin{cases} 1 \\ 0 \end{cases} \quad (5)$$

where $\mu_k \in \mathbb{R}$ is the measurement vector state matrix and $\mu_k(i)$ denotes the element $i$ in the state transmission matrix. Consequently, measurement data received under DoS attacks by the state estimator can be expressed in the following matrix:

$$\vec{z}_k = \begin{bmatrix} \mu_k(1) \times z_k \\ \mu_k(2) \times z_{k-1} \\ \dots \\ \mu_k(d+1) \times z_{k-d} \end{bmatrix} \quad (6)$$

where $\vec{z}_k$ is equal to the measurement data gathered from DoS attacks.

## IV. Proposed Method

To overcome some of the issues associated with existing approaches, in this section, we propose a generalized deep learning model that works with raw imbalanced datasets. The proposed model generates a new balanced representation from a raw dataset and feeds it to an ensemble deep learning model for classification. The deep learning model consists of multiple unsupervised SAE that learns new representations from imbalanced datasets. The SAE attack detection model utilizes multiple Autoencoders (AE) to extract a new representation from unlabeled data to obtain different patterns. Then, new representations from each SAE are passed to a DNN via super vector and concatenated using a fusion activation vector. Finally, a DT is used, as a binary classifier, to detect attacks from the newly merged representations. The schematic of the proposed model is presented in Figure 2.

### A. The Proposed Ensemble Deep Representation Learning Model

Most existing approaches proposed in literature neglect the fact that real ICSs are highly imbalanced (the number of attack samples is a lot less than the number of normal samples). This will result in a low f-measure, which reflects the low performance of these models in an imbalanced environment like ICSs, thereby makes them impractical for real-world use cases.

Once a model is directly trained with a highly imbalanced dataset, the new malicious data are likely to be misclassified. To address this problem, we propose an ensemble deep representation-learning model based on SAE to enhance the overall performance of the model. This is done through extracting an equal balanced set and passing it to multiple AE to generate new representations. The input sample $x_i$ in a sample set $X$ corresponding to the hidden layer is represented in the following equation:

$$h_i = f(x_i) = \sigma(W_1 x_i + b_1) \quad (7)$$

where W and b represent the weight matrix of neurons and bias vector of all neurons between the input and hidden layers, respectively [43]. $\sigma$ is a function of the hidden layer used after beginning the training process by updating the next input layer to construct a set of stacked multi-layer AEs. Although using an ensemble model has increased the computational efficiency by a little, it was evident that utilizing multiple AE would lead to much better f-measure scores.

To enhance the performance of each AE, a dropout layer is added to enhance the generalization of our model by reducing the reliance of the output on a specific set of parameters. Also, the number of nodes and layers was selected through cross-validation of various networks with critical analysis of loss history and validation accuracy. Binary Cross-Entropy (BCE) is used as the cost function, represented by:

$$J = -\frac{1}{N} \sum_{i=1}^{N} y_i . \log(p(y_i)) + (1 - y_i) . \log(1 - p(y_i)) \quad (8)$$

where $y_1$ and $y_2$ represent attack and normal samples, respectively. $N$ is the total number of samples, and $p(y)$ is the expected likelihood of an attack sample. BCE was used over Mean Squared Error (MSE) to prevent neuron weight changes in the hidden layer of the AE from getting smaller and smaller, thereby stalling out the system.

### B. The Proposed Ensemble Deep Learning Attack Detection Model

Once the new representations are generated form the imbalanced dataset, they are fed to an ensemble of DNN classifiers to detect normal from abnormal behaviors. The results from each DNN is then concatenated, via super vector using a fusion activation function, and passed on to a DT classifier to detect attacks from the newly merged representations. A DT classifier was selected based on multiple tests using different machine learning classifiers, with DT providing the best performance results. The fusion activation function of the sigmoid layer is represented by the following equation:

$$L_1 = \sum_{i=1}^{m} y_i \log(t_i) . w_s + (1 - y_i) \log(1 - t_i) . w_l \quad (9)$$

where $L_1$ is the fusion activation function of the sigmoid layer, $y_i$ is the label of $i - th$ sample, $t_i$ is the prediction of the $i$-th sample. $w_s$ and $w_l$ are weights of unstable and stable samples, respectively. $w_s$ is set larger than $w_l$ to improve the detection of unstable samples, and $w_l$ is always set to 1 as a benchmark to mine unstable patterns effectively [33].

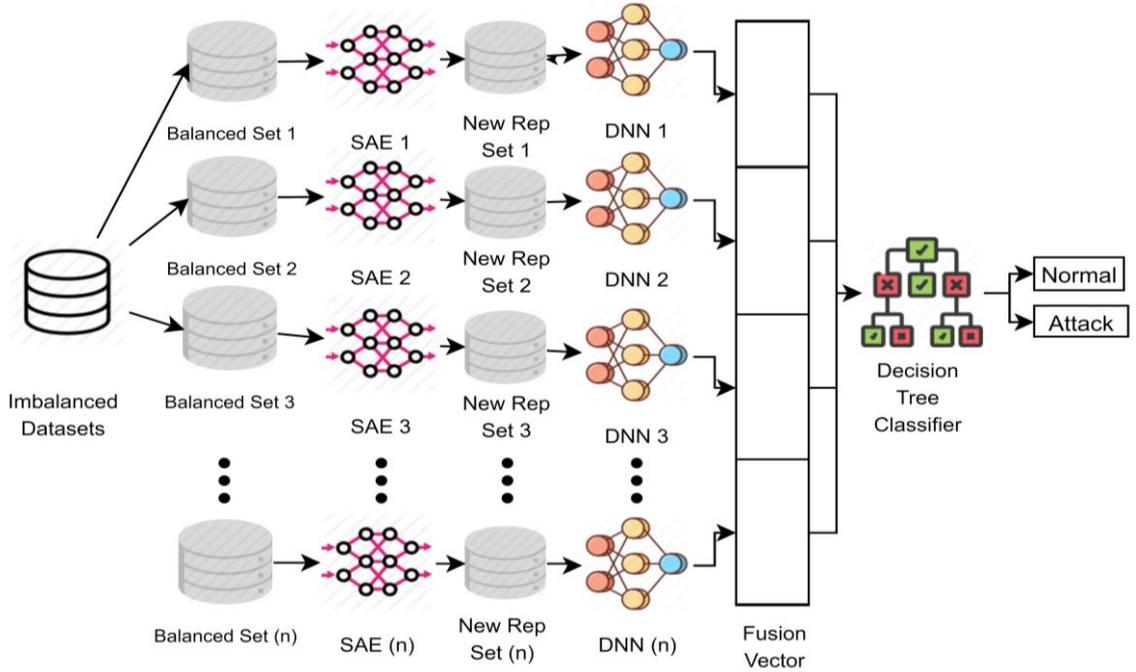

**FIGURE 2.** Overview Model of Stacked Autoencoder Algorithm

The AEs were tested in a for loop, using a different number of layers, neurons, batch sizes, loss and activation functions, optimizers, epochs, and dropout layers, to achieve better accuracy and f-measure. Both SAE and DNN utilize BCE cost function as well as Rectified Linear Unit (ReLU) activation function to achieve best performance measures, represented by:

$$ReLU(x) = \max(0, x) \quad (9)$$

where $x$ is the observation.

The pseudocode of the proposed attack detection algorithm is shown in Algorithm 1.

## IV. Case Studies and Results Analysis

### A. Data Preparation
Ideally, using new real SCADA data should be appraised, but due to the limitations of available real datasets, this study resorted to realistic ICS datasets obtained in 2015 and 2018. In this section, two different ICS datasets are used to evaluate the performance of the proposed algorism's efficiency against random ICS models.

- **Gas Pipeline (GP):** This dataset is obtained from a gas pipeline system and contains a Modbus validation frame of a preprocessed dataset in an Attribute-Relation File Format (ARFF) to help researchers use specialized preprocessing techniques. It also has a deep packet inspection of the Modbus frame with each line representing one network transaction. The dataset contains 17 features, with a total of 274628 observations split into 219702 (80%) samples for training and 54925 (20%) for testing [44].

- **Secure Water Treatment (SWaT):** This dataset includes 11 days of continuous operation, in which 7 days were recorded under normal operation conditions and 4 days with attack scenarios. SWaT contains a total of 51 features, collected from network traffic ports, sensors, and actuators, with a total of 1048576 observations split into 838860 (80%) samples for training and 209715 (20%) for testing [45].

### B. Evaluation Metrics
When it comes to the security of ICSs, the concern revolves around detecting cyber-attacks while achieving high f1-scores on imbalanced datasets, thereby minimizing the rate of false alarms. As with standard machine learning benchmarking metrics, this work considers True Positives (TP), False Positives (FP), True Negatives (TN), and False Negatives (FN), defined in Table I, as the performance evaluation metrics for the attack detection models.

TABLE I
UNITS FOR MAGNETIC PROPERTIES

|  | Attacked | Secure |
|---|---|---|
| Classified as Attacked | TP | FP |
| Classified as Secure | FN | TN |

**Algorithm 1:** The proposed ensemble attack detection SAE model

**Data:** Input all datasets including **Normal** and **Attack** samples
**Training Phase:**
**for** 10 folds of cross-validation **do**
   Split the dataset into **Training (80%)** and **Testing (20%)** sets
   Normalized the data: $z = \frac{x - min(x)}{max(x) - min(x)}$
   Separate the samples into four balanced sets with each containing **(50 % Normal, 50% Attack)** samples.
   **Training the SAE model**
   Feed each balanced set to the SAE model to generate new representations of data
   **for** *number of epochs* **do**
      **for** *number of batches in the balanced set 1* **do**
         Train the autoencoder: min $\mathcal{L}(x_i, \hat{x}_i)$
         **Loss function:** Binary Cross Entropy (BCE),
         **Optimizer:** Adam
      **end**
      **for** *number of batches in the balanced set 2* **do**
         Train the autoencoder: min $\mathcal{L}(x_i, \hat{x}_i)$
         **Loss function:** BCE, **Optimizer:** Adam
      **end**
      **for** *number of batches in the balanced set 3* **do**
         Train the autoencoder: min $\mathcal{L}(x_i, \hat{x}_i)$
         **Loss function:** BCE, **Optimizer:** Adam
      **end**
      **for** *number of batches in the balanced set 4* **do**
         Train the autoencoder: min $\mathcal{L}(x_i, \hat{x}_i)$
         **Loss function:** BCE, **Optimizer:** Adam
      **end**
   The new representations sets are then used to train four DNN models for anomaly detection
   **Training the ensemble DNN detection model:**
   Train 4 DNN models, each corresponding to the 4 new representation sets
   **for** *number of estimators* **do**
      Train the DNN model on set 1
      **Loss function:** BCE, **Optimizer:** Adam
   **end**
   **for** *number of estimators* **do**
      Train the DNN model on set 2
      **Loss function:** BCE, **Optimizer:** Adam
   **end**
   **for** *number of estimators* **do**
      Train the DNN model on set 3
      **Loss function:** BCE, **Optimizer:** Adam
   **end**
   **for** *number of estimators* **do**
      Train the DNN model on set 4
      **Loss function:** BCE, **Optimizer:** Adam
   **end**
**end**
**Fusion Layer:**
Merge the new representations from each DNN to form a **Super-vector** using the **NumPy** concatenating function
Pass the super vector to a final DT binary detection model
**Training the DT model:**
**for** *number of estimators* **do**
   Train DT classifier on new merged representation
**end**
**Testing Phase:**
Normalize the test sample
Pass 4 test sets through the SAEs
Pass each new generated representation from SAE to DNN for anomaly detection
Fuse the output of the DNN into a super vector
Pass the super-vector to a DT model for binary classification
**Output:** Normal/Attack label

The performance of the machine learning algorithms is measured by the following metrics [44]:

- **Accuracy:** Ratio of samples classified correctly over the entire dataset.

$$Acc = \frac{TP + TN}{TP + TN + FP + FN} \quad (10)$$

- **Precision:** The percentage of correctly classified positive samples.

$$Prec = \frac{TP}{TP + FP} \quad (11)$$

- **Recall:** The ratio of correctly predicted positive samples over the total samples of the corresponding class.

$$Rec = \frac{TP}{TP + FN} \quad (12)$$

- **F1 Score:** Harmonic mean of precision and recall.

$$F1 = \frac{2 \times TP}{2 \times TP + FN + FP} \quad (13)$$

The F1-score aims to find an equal balance between precision and recall, which is highly important in performance evaluation for imbalanced datasets (i.e., the number of attack samples are a lot less than the number of normal samples).

*C. Performance Analysis*

***General Performance Analysis-*** In this section, two different ICS datasets gathered from a gas pipeline system and a water treatment facility were used to evaluate the performance of the proposed method. Results were compared with DNN, RF, DT, and Adaboost based classifiers along with multiple peer approaches in the current literature. Tables II and III provide a summary of performance evaluation metrics results, including accuracy, precision, recall, and F1-score. As illustrated, the results of the proposed method, in both datasets, outperform existing techniques in all four metrics, and most importantly on f-measure, which highlight the efficiency of the proposed model in imbalanced ICS environments.

***Imbalanced Testing-*** To evaluate the efficiency of the proposed method under different imbalanced conditions, we have tested the model with different imbalanced ratios. Imbalanced ratio of 0.1 means 10% of the attack samples were used, and in the same way, an imbalanced ratio of 1 means a %100 is utilized.

TABLE II
SUMMARY OF THE RESULTS AND PERFORMANCE COMPARISON ON THE GAS PIPELINE DATASETS

| Method | Acc | Pre | Rec | F1 |
|---|---|---|---|---|
| **Proposed** * | **0.96** | **0.9463** | **0.9372** | **0.9383** |
| CAE [26] | 0.86 | 0.8806 | 0.8612 | 0.8358 |
| SVM [27] | 0.92 | 0.7820 | 0.9360 | 0.8520 |
| LSTM [30] | 0.92 | 0.9400 | 0.7800 | 0.8500 |
| NB [36] | 0.90 | 0.8195 | 0.7692 | 0.8595 |
| DT | 0.86 | 0.9159 | 0.6808 | 0.7239 |
| DNN | 0.84 | 0.8994 | 0.6389 | 0.6709 |
| RF | 0.83 | 0.9142 | 0.6298 | 0.6591 |

TABLE III
SUMMARY OF THE RESULTS AND PERFORMANCE COMPARISON ON THE SWaT DATASETS

| Method | Acc | Pre | Rec | F1 |
|---|---|---|---|---|
| **Proposed** | **99.67** | **0.97** | **0.99** | **0.99** |
| SVM [37] | - | 0.93 | 0.699 | 0.79 |
| RNN [37] | - | 0.94 | 0.699 | 0.8 |
| ID CNN [29] | - | 0.96 | 0.799 | 0.87 |
| TABOR [35] | 94.99 | 0.86 | 0.79 | 0.82 |
| AE [38] | - | 0.89 | 0.80 | 0.84 |
| AE Frequency [38] | - | 0.92 | 0.83 | 0.87 |
| DNN | 96.24 | 0.96 | 0.95 | 0.95 |

As shown in Figures 3-6, results of the proposed method exceed other techniques with a flat curve in all metrics for the GP dataset. This verifies the robustness of the proposed method as its performance is not affected by different imbalanced ratios. Although other methods have an acceptable accuracy, the recall and precision are significantly lower than that of the proposed method. However, our proposed method maintains consistent results in all four metrics.

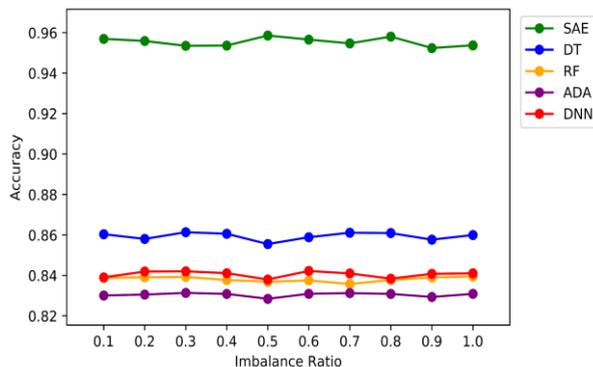

**FIGURE. 3** Accuracy under different imbalance ratios for the Gas Pipeline dataset.

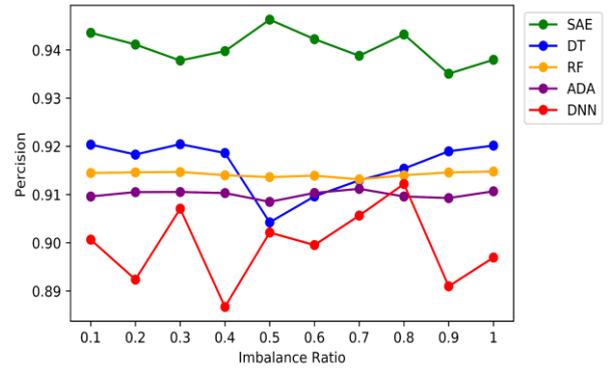

**FIGURE. 4** Precision under different imbalance ratio for Gas Pipeline dataset.

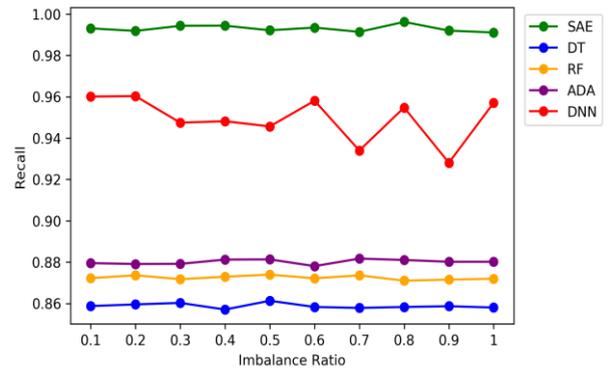

**FIGURE. 5** Recall under different imbalance ratios for the Gas Pipeline dataset.

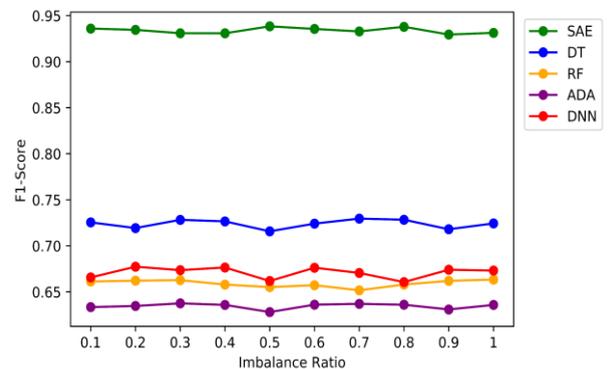

**FIGURE. 6** F1-score under different imbalance ratios for the Gas Pipeline dataset.

For further analysis, the proposed model was evaluated on the SWaT dataset, too. Since the model is generalized for different ICS environment, the proposed model was tested without any modification on the model structure or parameters. As illustrated in Figure 7-10, the proposed method outperforms existing techniques in all four metrics. Better performance compared to the first case study could be attributed to the fact that there are more samples for training in the SWaT dataset than what exists in the GP dataset.

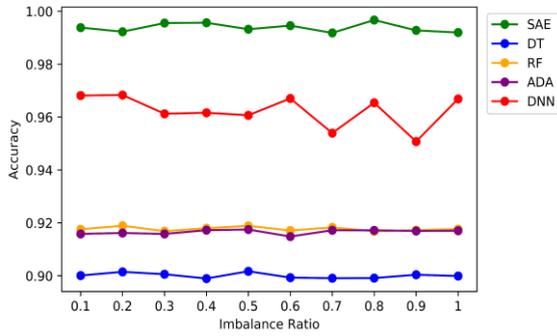

**FIGURE. 7** Accuracy under different imbalance ratios for the SWaT dataset.

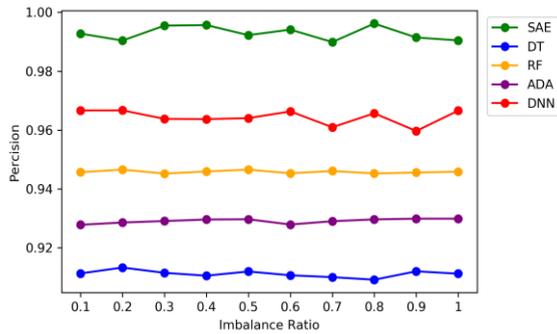

**FIGURE. 8** Precision under different imbalanced ratios for the SWaT dataset.

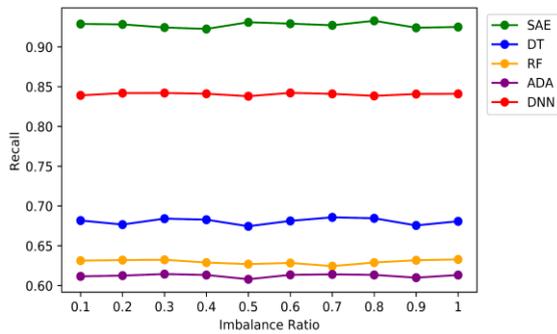

**FIGURE. 9** Recall under different imbalanced ratios for the SWaT dataset.

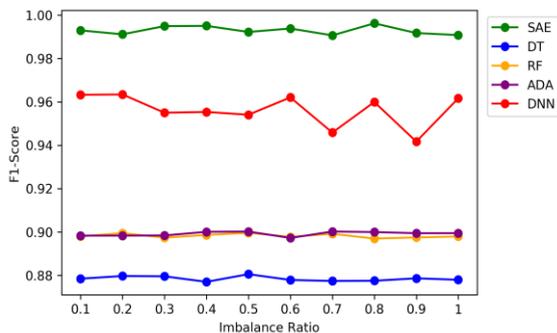

**FIGURE. 10** F1-score under different imbalance ratios for the SWaT dataset.

## V. CONCLUSION

Critical infrastructures are complex cyber and physical systems that structure the lifeline of modern society, and their reliable and secure operations are essential to national security. In this paper, we proposed a generalized ensemble deep learning-based cyber-attack detection method specifically designed for ICS. The proposed technique includes a deep representation-learning model, which constructs new balanced representations from the raw imbalanced dataset. The new representations are then used in an ensemble deep learning algorithm based on DNN and DT classifiers to detect cyber-attacks. The performance of the proposed model is verified using two different ICS datasets obtained from real critical infrastructure facilities. Our proposed approach outperformed conventional classifiers with %10 higher f1-score in both datasets evaluated and produced higher accuracy, with %95.86 for the Gas Pipeline dataset and %99.67 for the Secure Water Treatment dataset. Results were compared with traditional classifiers, such as RF, DNN, and ADA, along with multiple peer proposed approaches in the current literature. The proposed approach outperformed other techniques in all four-evaluation metrics. Although our approach performed better than existing techniques, there is room for improvement when dealing with few samples, as illustrated in the GP dataset. Additionally, identifying the attack type and its location is also very important to prevent processing downtime and computation efficiency once an attack is detected. Therefore, our future work will focus on optimizing the accuracy of the proposed method and developing an additional model to identify different attack types and their locations. This will avoid critical system failure and improve the network security of ICSs against similar cyber-attacks.

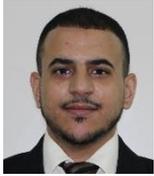
**Abdulrahman Al-Abassi** received his B.S. degree in Engineering System, specializing in the mechanics field in 2018 from the University of Guelph. Following that, he worked at an industrial facility as an operations process technician for almost a year. Currently, he is an M.A.S.c student in the Engineering Systems and Computing group at the University of Guelph. His current interests include defining deep learning methods for detecting, identifying and locating cyber-attacks in Industrial Control Systems.

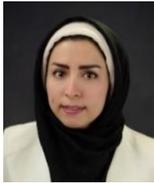
**HADIS KARIMIPOUR** received a Ph.D. degree in Electrical Engineering from the University of Alberta in Feb. 2016. Before joining the University of Guelph, she was a postdoctoral fellow at the University of Calgary working on anomaly detection and security analysis of the smart power grids. She is currently an Assistant Professor at the School of Engineering at the University of Guelph, Guelph, Ontario. Her research interests include power system analysis, cyber-physical modeling, cyber-security of the smart grids, and parallel and distributed computing. She is a member of the IEEE and IEEE Computer Society. She serves as the Chair of the IEEE Women in Engineering (WIE) and chapter chair of IEEE Information Theory in the Kitchener-Waterloo section.

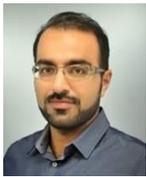
**ALI DEHGHANTANHA** is the director of the Cyber Science Lab at the University of Guelph, Ontario, Canada. His lab is focused on building AI-powered solutions to support cyber threat attribution, cyber threat hunting and digital forensics tasks in the Internet of Things (IoT), Industrial IoT, and Internet of Military of Things (IoMT) environments. Ali has served for more than a decade in a variety of industrial and academic positions with leading players in Cyber-Security and Artificial Intelligence. Prior to joining UofG, he has served as a Sr. Lecturer in the University of Sheffield, UK and as an EU Marie-Curie International Incoming Fellow at the University of Salford, UK. He has Ph.D. in Security in Computing and a number of professional certifications including CISSP and CISM.

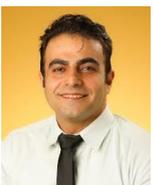
**REZA M. PARIZI** received the B.Sc. and M.Sc. degrees in Computer Science in 2005 and 2008, respectively, and the Ph.D. degree in software engineering in 2012. He is the director of Decentralized Science Lab (dSL) at Kennesaw State University, GA, USA. He is a consummate AI technologist and software security researcher with an entrepreneurial spirit. He is a senior member of the IEEE, IEEE Blockchain Community, and ACM. Prior to joining KSU, he was a faculty at the New York Institute of Technology. His research interests are R\&D in decentralized AI, cybersecurity, blockchain systems, smart contracts, and emerging issues in the practice of secure software-run world applications.